\documentclass[prd,twocolumn,superscriptaddress,altaffilletter,amssymb,showpacs,nofootinbib]{revtex4}
\usepackage[dvips]{graphicx}

\usepackage{amsmath}
\newcommand{\be}{\begin{equation}}
\newcommand{\ee}{\end{equation}}
\newcommand{\bea}{\begin{eqnarray}}
\newcommand{\eea}{\end{eqnarray}}

\begin{document}

\title{Dynamics of FRW Universes Sourced by Non-Linear Electrodynamics}

\author{Ricardo Garc\'{\i}a-Salcedo}\email{rigarcias@ipn.mx}
\affiliation{Centro de Investigacion en Ciencia Aplicada y Tecnologia Avanzada - Legaria
del IPN, M\'{e}xico D.F., M\'{e}xico.}

\author{Tame Gonzalez}\email{tame@uclv.edu.cu}
\affiliation{Departamento de F\'{\i}sica, Universidad Central de Las Villas, 54830 Santa
Clara, Cuba.}

\author{Claudia Moreno}\email{claudia.moreno@cucei.udg.mx}
\affiliation{Departamento de F\'{\i}sica y Matem\'aticas, Centro Universitario de
Ciencias Ex\'actas e Ingenier\'{\i}as, Corregidora 500 S.R., Universidad de
Guadalajara, 44420 Guadalajara, Jalisco, M\'exico}\affiliation{Part of the Instituto Avanzado de Cosmolog\'ia (IAC) collaboration http://www.iac.edu.mx/}

\author{Israel Quiros}\email{israel@uclv.edu.cu}
\affiliation{Departamento de F\'{\i}sica, Universidad Central de Las Villas, 54830 Santa
Clara, Cuba.}\affiliation{Part of the Instituto Avanzado de Cosmolog\'ia (IAC) collaboration http://www.iac.edu.mx/}
\date{\today }

\begin{abstract}
We apply the dynamical systems tools to study the (linear) dynamics of Friedmann-Robertson-Walker universes that are fuelled by non-linear electrodynamics. We focus, mainly, in two particular models. In the first model the cosmic evolution is fuelled by cold dark matter, a cosmological constant and a non-linear electrodynamics field. In the second case non-singular cosmology and late-time accelerated expansion are unified in a model where the Einstein's field equations are sourced only by cold dark matter and a non-linear electrodynamics field. It is shown that, in contrast to previous claims, the cosmological effects coming from the non-linear electrodynamics field are not as generic as though. In fact, critical points in the phase space that could be associated with non-linear electrodynamic effects are not found.
\end{abstract}

\maketitle

\section{Introduction}

Studying the equations of the non-linear electrodynamics (NLED) is an attractive subject of research in general relativity (GR) thanks to the fact that such quantum phenomena as vacuum polarization can be implemented in a classical model through their impact on the properties of the background space-time. Exact solutions of the Einstein's field equations coupled to NLED may hint at the relevance of the non-linear effects in strong gravitational and magnetic fields. It has been speculated, in particular, that very strong electromagnetic fields might help avoiding the occurrence of space-time singularities \cite{27}. The impact of very strong electromagnetic fields (and of the NLED effects) regarding the causality issue in cosmology is also of relevance \cite{Novello0}.

A different subject of research within the cosmological setting in GR coupled to NLED, is related with the chance for the NLED field to fuel primordial inflation. The cosmological inflationary scenario was proposed for the first time in the Ref. \cite{20} where, in order to solve (or avoid) several problems of the standard model -- such as the flatness and the horizon problems, amongts others --, it was anticipated that a self-interacting scalar field with a particular form of the self-interaction potential as the source of the Einstein's field equations, might cause the universe to expand in an inflationary (super-accelerated) fashion. The inflationary paradigm has been supported by the observational evidence \cite{21}.

In the reference \cite{15} an anisotropic cosmological model coupled to Born-Infeld NLED was explored. It was found that this model might explain early time inflation \cite{28}. Thanks to the inflationary stage, the initially anisotropic universe might eventually isotropize. An alternative non-isotropic model where, additionally to the Born-Infeld NLED field, a cosmological constant was added to the Einstien's field equations, has been studied in \cite{Vollick0}. Yang-Mills cosmology with a non-Abelian Born-Infeld action has been also considered \cite{17}.

Magnetic universes (vanishing electric component) have been investigated within the context of theories given by the Lagrangian density ${\cal L}_1=-\frac{1}{4}F+\alpha F^2+\beta G^2$ (see below) \cite{16}. The non-linear term $\propto F^2$ may cause the universe to bounce thus avoiding the initial (big-bang) singularity. Then, in Ref.\cite{Vollick1} it was demonstrated that the inclusion of a non-vanishing electric component $E$ where $E^2\simeq B^2$, removes the bounce and, in consequence, the universe starts its evolution in a singular state. Models with Lagrangian density of the form ${\cal L}_2=-\frac{1}{4}F-\frac{\gamma}{F}$ may account for the late-time stage of accelerated expansion of magnetic universes \cite{23}. In the later case, if a non-vanishing electric component $E$ is considered, accelerated expansion is allowed only when $E^2<3B^2$ \cite{Vollick1}.

Aim of the present investigation is to explore the asymptotic properties of cosmological models where general relativity is coupled to NLED given by the above Lagrangian densities ${\cal L}_1$ and ${\cal L}_2$. We will rely on the use of the dynamical systems tools with the hope that such relevant concepts as past and/or future attractors, or saddle equilibrium points, could be correlated with generic cosmological behaviour. Our goal will be to write the cosmological (Einstein's) equations in the form of an autonomous system of ordinary differential equations (ODE). Although these equations are non-linear, we will expand them in the neighbourhood of the equilibrium (critical) points up to the linear approximation. Evaluating the sign of the real parts of the eigenvalues of the corresponding linearization matrices will allow us to judge about the stability of the critical points. We will show, in particular, that equilibrium points corresponding to dominance of NLED effects are not found, signaling at a fundamental drawback of the models. Here we use natural units ($8\pi G=8\pi/m_{Pl}^2=\hbar=c=1$).

\section{$\Lambda$-CDM-NLED Model}

In this and in the next sections we shall focus in a FRW cosmological model with flat spatial sections, fuelled by three generic sources: i) pressureless cold dark matter (CDM), ii) dark energy in the form of a non-negative cosmological constant $\Lambda\geq 0$, and iii) non-linear electrodynamics (NLED). While the first and second components (CDM and $\Lambda$-dark energy, respectively) are required for consistency of the late-time cosmological evolution, the third component (properly the non-linear electrodynamics field) is important at early as well as at intermediate times. At early times departure from standard radiation-like behaviour is apparent, however, later on in the course of the cosmic evolution the NLED field behaves just like a standard radiation field. 

In the next section we want to explore the asymptotic properties of the above model that could be correlated with generic cosmological behaviour. This will allow us to make a judgment about the actual relevance of NLED on early-time dynamics, in particular for the bounce of the universe. Later on, in section IV, we shall explore a model where the bounce and late-time inflationary dynamics are originated by a single NLED field. 

In general the Lagrangian density of NLED is given by ${\cal L}={\cal L}(F,G)$ where $F=F_{\mu \nu }F^{\mu \nu }$ and $G=F^{\mu \nu }\check{F}_{\mu \nu }$ are the electromagnetic invariants $\left(\check{F}_{\mu \nu }=\frac{1}{2}\epsilon_{\mu \nu \alpha \beta }F^{\alpha \beta }\right)$. The energy-momentum tensor that is associated with this Lagrangian can be written in the following way: 

\be T_{\mu\nu}=-4(\partial _F {\cal L}) F_\mu^\alpha F_{\alpha\nu}+(G\partial_G {\cal L}-{\cal L})g_{\mu\nu}.\label{Tmn}\ee 

In order to meet the requirements of homogeneous and isotropic cosmologies (as, in particular, the one associated with FRW space-times), the energy density and the pressure of the NLED field should be evaluated by averaging over volume (for details see \cite{novellocqg} and references therein). Additionally it has to be assumed that the electric and magnetic fields, being random fields, have coherent lengths that are much shorter than the cosmological horizon scales. After several considerations the energy-momentum tensor of the electro-magnetic (EM) field associated with the Lagrangian density ${\cal L}={\cal L}(F,G)$ can be written in the form of the energy-momentum tensor for a perfect fluid:

\be T_{\mu\nu}=\left(\rho+p\right)u_{\mu}u_{\nu}-pg_{\mu\nu}, \label{Tpf}
\ee where 

\bea &&\rho=-{\cal L}+G\partial_G {\cal L}-4\partial_F {\cal L} E^{2},\label{rho}\\
&&p={\cal L}-G\partial_G {\cal L}-\frac{4}{3}\left(2B^2-E^2\right)\partial_F {\cal L},\label{p}\eea $E$ and $B$ being the averaged electric and magnetic fields respectively. For the purposes of the present investigation we shall consider the following Lagrangian density that has been formerly studied in \cite{16}:
 
\be {\cal L}=-\frac{1}{4}F+\alpha F^{2}+\beta G^{2},\label{lag}\ee where $\alpha $ and $\beta $ are arbitrary constants. As mentioned in \cite{novellocqg}, a particularly interesting case arises when only the average of the magnetic part $B$ is different from zero, leading to the so called magnetic universe. This case turns out to be relevant in cosmology as long as the averaged electric field $E$ is screened by the charged primordial plasma, while the magnetic field lines are frozen \cite{lemoine}. Let us focus here, exclusively, in a non-singular universe proposed in Ref.\cite{lemoine} (Lagrangian density (\ref{lag}) with $\beta=0$). According to this scenario the energy density (\ref{rho}) and effective presure (\ref{p}) are given by:

\bea &&\rho_B=\frac{1}{2}B^2\left(1-8\alpha B^2\right),\label{rho1}\\
&&p_B=\frac{1}{6}B^2\left(1-40\alpha B^2\right),\label{p1}\eea respectively, where, in order to produce the bounce, the magnetic field $B$ has to be associated with the following evolution law: $B=\frac{B_0}{a^2}$ ($B_{0}$ is a constant of integration). 

Notice that the energy density of the NLED field $B$ vanishes at the maximum allowed value of the magnetic field $B=1/\sqrt{8\alpha}$. $\rho_B$ is a minimum at $B=0$, while at $B=1/\sqrt{16\alpha}$ it is a maximum instead. At the maximum the following relationship between the energy density and the parametric pressure arises: $p_B=-\rho_B$, i. e., at the maximum the NLED field behaves as vacuum does. In consequence the equation of state parameter $\omega_B$ fluctuates between $1/3$ (radiation) and $-1$ (vacuum energy). This means, in turn, that the NLED field could contribute towards primordial (early-time) inflation. However, it will be shown here that this inflationary behaviour is not as generic as though since it can not be correlated with any equilibrium point in the phase space of the model. Notice that the energy density is positive $(\rho_B>0)$ whenever

\be B<\frac{1}{2\sqrt{2\alpha}},\nonumber\ee meanwhile the pressure will be negative only if 

\be B>\frac{1}{2\sqrt{10\alpha}}.\nonumber\ee The strong energy condition is violated ($\rho_B+3p_B<0$) whenever

\be B>\frac{1}{2\sqrt{6\alpha}}.\nonumber\ee In the later case the NLED field might fuel early-time inflationary dynamics. As already said, it will be demonstrated that inflationary NLED does not represent a critical point in the phase space of the model, meaning that it is unlikely that a NLED field might be a generic source of cosmic inflation.

The starting point for the present investigation will be the following Einstein's field equations (as usual $H=\frac{\dot{a}}{a}$ is the Hubble parameter, the overdot accounts for derivative in respect to the cosmic time $t$, while $\rho_B$ and $p_B$ are given by (\ref{rho1}) and (\ref{p1}) respectively):

\bea &&3H^2=\rho_{CDM}+\Lambda+\rho_B,\nonumber\\
&&2\dot H=-\rho_{CDM}-\left(\rho_B+p_B\right),\label{feqs}\eea plus the following conservation equations:

\be \dot\rho_{CDM}+3H\rho_{CDM}=0,\label{contcdm}\ee and the continuity equation for the NLED magnetic field,
 
\be \dot\rho_B+3H(\rho_B+p_B)=0.\label{cont}\ee Equations (\ref{feqs}), (\ref{contcdm}) and (\ref{cont}) form the mathematical basis of the model of interest in the present paper.

\section{Dynamical Systems Study}

The dynamical systems tools offer a very useful approach to the study of the asymptotic properties of the cosmological models \cite{coley}. In order to be able to apply these tools one has to (unavoidably) follow the steps enumerated below.

\begin{itemize}

\item First: to identify the phase space variables that allow writing the system of cosmological equations in the form of an autonomus system of ordinary differential equations (ODE). There can be several different possible choices, however, not all of them allow for the minimum possible dimensionality of the phase space.

\item Next: with the help of the chosen phase space variables, to build an autonomous system of ODE out of the original system of cosmological equations. 

\item Finally (some times a forgotten or under-appreciated step): to indentify the phase space spanned by the chosen variables, that is relevant to the cosmological model under study.

\end{itemize}

After this one is ready to apply the standard tools of the (linear) dynamical systems analysis.

\subsection{The Autonomous System of ODE}

We introduce the following phase variables:

\be x\equiv\frac{B}{\sqrt{6}H},\;\;y\equiv\frac{\sqrt{\Lambda}}{\sqrt{3}H}
,\;\; v\equiv 4\sqrt\alpha H.\ee We will be focused, exclusively, on expanding FRW
universes, so that $x\geq 0$, $y\geq 0$, and $v\geq 0$ ($B\geq 0$). 

After the above choice of variables of the phase space, the system of cosmological equations (\ref{feqs},\ref{contcdm},\ref{cont}) can be translated into the following autonomous system of ordinary differential equations (ODE):

\bea &&x'=-x\left(2+\frac{H'}{H}\right),\nonumber\\
&&y'=-y\;\frac{H'}{H},\;v'=v\;\frac{H'}{H},\nonumber\\
&&\frac{H'}{H}=-\frac{3}{2}[1-y^2+\frac{x^2}{3}(1-15x^2v^2)],\label{ode}\eea where the prime denotes derivative with respect to the new time variable $\tau\equiv\ln a$ -- properly the number of efoldings. The phase space where to look for the equilibrium points of the above system of ODE is the following:

\bea &&\Psi=\{(x,y,v): 0\leq y^2+x^2(1-3x^2v^2)\leq 1,\nonumber\\
&&\;\;\;\;\;\;\;\;\;\;\;\;\;\;x\geq 0,\;y\geq 0,\;v\geq 0,\;x^2v^2\leq 1/3\}.\label{phasespace}\eea

Additionaly it will be helpful to have the following parameters of observational importance $\Omega_B=\rho_B/3H^2$ -- the NLED field dimensionless energy density
parameter --, and the equation of state (EOS) parameter $\omega_B=p_B/\rho_B$, written in terms of the phase space variables:

\be \Omega_B=x^2(1-3x^2v^2),\; \ee \be \omega_B=\frac{1}{3}\left(\frac{1-15x^2v^2}{1-3x^2v^2}\right), \ee

\be \Omega_{\Lambda}=y^2, \hspace{1cm} \Omega_{CDM}=1-\Omega_{\Lambda}-\Omega_B . \ee

Notice also, that for the deceleration parameter one has $q=-(1+H'/H)$.

\begin{table*}[tbp]
\caption[crit]{Properties of the critical points for the autonomous system (\ref{ode}), and eigenvalues of the linearization matrices.}
\label{tab1}
\begin{tabular}{@{\hspace{4pt}}c@{\hspace{14pt}}c@{\hspace{14pt}}c@{\hspace{14pt}}c@{\hspace{14pt}}c@{\hspace{14pt}}c@{\hspace{14pt}}c@{\hspace{14pt}}c@{\hspace{14pt}}c@{\hspace{14pt}}c@{\hspace{14pt}}c@{\hspace{14pt}}c}
\hline\hline\\[-0.3cm] 
$P_{i}$ & $x$ & $y$ & $v$ & $\Omega_{CDM}$ & $\Omega_\Lambda$& $\Omega_B$ & $\omega_B$ & $q$ & $\lambda_1$ & $\lambda_2$ & $\lambda_3$\\[0.1cm] \hline\\[-0.2cm] 
$P_1$ & $0$ & $0$ & $0$ & $1$ & $0$ & $0$ & $1/3$ & $1/2$  & $-3/2$ & $-1/2$ & $3/2$ \\[0.2cm] 
$P_2$ & $1$ & $0$ & $0$ & $0$ & $0$ & $1$ & $1/3$ & $1$  & $-2$ & $1$ & $2$ \\[0.2cm] 
$P_3$ & $0$ & $1$ & $4\sqrt{\alpha\Lambda/3}$ & $0$ & $1$ & $0$ & $1/3$ & $-1$ & $-3$ & $-2$ & $0$\\[0.2cm] 
\hline\hline\end{tabular}
\end{table*}

\begin{figure}[ht!]
\begin{center}
\includegraphics[width=8cm,height=6.5cm]{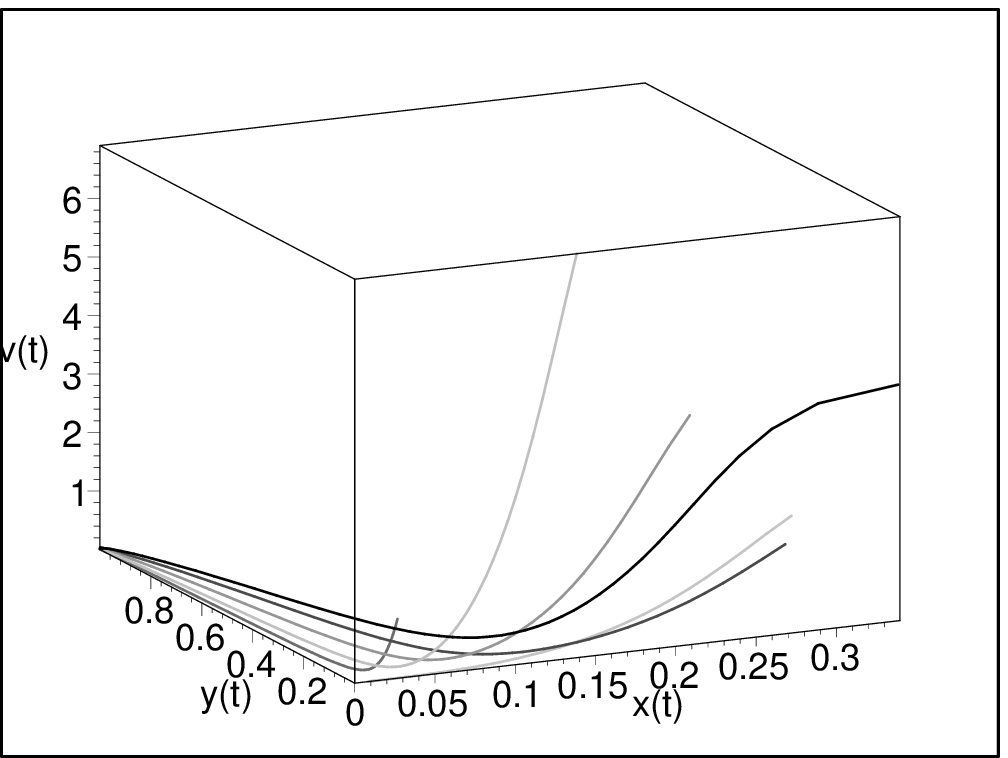}
\includegraphics[width=4cm,height=4cm]{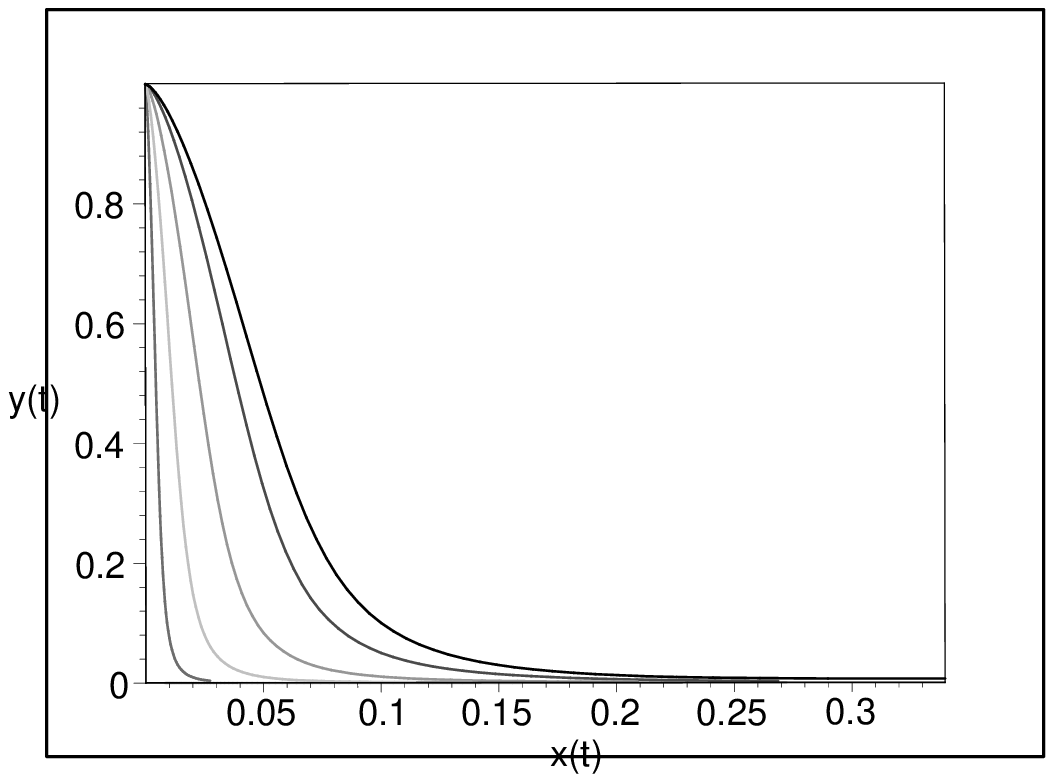}
\includegraphics[width=4cm,height=4cm]{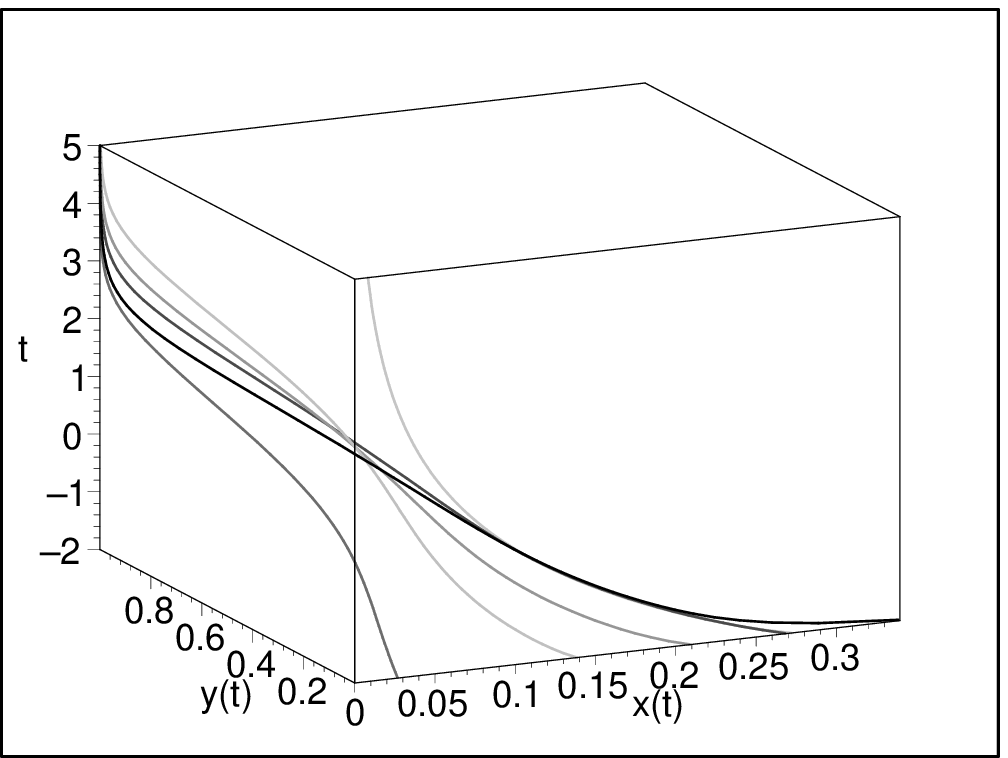}
\vspace{0.3cm}
\caption{The orbits of the autonomous system of ODE (\ref{ode}) for arbitraryly chosen sets of initial conditions (upper panel). Only expanding cosmologies are reflected in the phase space. The projection of the corresponding orbits in the phase plane $(x,y)$ are shown in the left hand lower panel, while the flux in time of the orbits corresponding to expansion are shown in the right hand lower panel. Notice that the de Sitter solution $(0,1,4\sqrt{\alpha\Lambda/3})$ is the late-time (future) attractor of these orbits, while the solution dominated by the CDM $(0,0,0)$ is always a saddle critical point in the phase space.}\label{fig1}
\end{center}
\end{figure}

\subsection{Critical Points}

The critical points of the autonomous system of equations (\ref{ode}) and the eigenvalues of the corresponding linearization (Jacobian) matrices, are summarized in table \ref{tab1}.

There are three equilibrium points at all: i) CDM-dominated solution (point $P_1$ in Tab.\ref{tab1}), ii) radiation-dominated solution (point $P_2$), and iii) de Sitter -- inflationary -- solution (point $P_3$). The latter is a non-hyperbolic critical point so that we are not able to make final jugment about its stability. Maximum we can say is that there is a stable subspace attached to $P_3$, that is spanned by eigenvectors associated with $\lambda_1$ and $\lambda_2$. However, the plots of the orbits of the autonomous system of ODE (\ref{ode}) shown in the figure Fig.\ref{fig1}, demonstrate that the de Sitter solution is the late-time (future) attractor in the phase space of the model. The fact that, at the attractor point $v=4\sqrt{\alpha\Lambda/3}\rightarrow 0$, means that the non-linear electrodynamic effects -- represented by the constant parameter $\alpha$ -- have to be very weak. The remaining points $P_1$ and $P_2$ are always saddle critical points in $\Psi$. There is not any past attractor for the orbits of the system of ODE (\ref{ode}). From the cosmological point of view the model is viable since there are critical points associated with radiation and CDM dominance, as well as with the present stage of accelerated expansion. The same results would arise if one replaced the non-linear electrodynamics field by a standard Maxwell one. 

It comes as a surprise that the non-linear electrodynamics effects -- representing the strong field limit where the scalar of curvature is small and the volume of the universe attains its minimum \cite{novello2} --, can not be associated with any equilibrium point in the phase space $\Psi$ (\ref{phasespace}). This means that the NLED effects are not as generic as previously though. In particular, the non-singular behaviour correlated with the bounce, is not a generic property of the model where the radiation field is replaced by a (magnetic) NLED field. The bounce is just a particular solution of the model that, depending on the initial data chosen, might or might not occur, i. e., it could be only a unstable state in the phase space. 

In the next section we shall explore a model where also the fate of the cosmic expansion is determined by a NLED field (no $\Lambda$-term at all). We shall show, in particular, that from the cosmological point of view this model is not attractive, since, in this case there are no critical points in the corresponding phase space that could be identified with late-time cosmological dynamics.

\begin{table*}[tbp]
\caption[crit]{Properties of the critical points for the autonomous system (\ref{ode1}), and eigenvalues of the linearization matrices.}
\label{tab1'}
\begin{tabular}{@{\hspace{4pt}}c@{\hspace{14pt}}c@{\hspace{14pt}}c@{\hspace{14pt}}c@{\hspace{14pt}}c@{\hspace{14pt}}c@{\hspace{14pt}}c@{\hspace{14pt}}c@{\hspace{14pt}}c@{\hspace{14pt}}c@{\hspace{14pt}}c}
\hline\hline\\[-0.3cm] 
$P_{i}$ & $x$ & $v$ & $w$ & $\Omega_{CDM}$ & $\Omega_B$ & $\omega_B$ & $q$ & $\lambda_1$ & $\lambda_2$ & $\lambda_3$ \\[0.1cm] \hline\\[-0.2cm] 
$P_1$ & $0$ & $0$ & $0$ & $1$ & $0$ & $1/3$ & $1/2$ & $-3/2$ & $-1/2$ & $3$\\[0.2cm] 
$P_2$ & $1$ & $0$ & $0$ & $0$ & $1$ & $1/3$ & $1$ & $-2$ & $1$ & $4$ \\[0.2cm] 
\hline\hline\end{tabular}
\end{table*}

\section{Unified Description of Bouncing Cosmology Followed by Late-Time Accelerated Expansion}

In this section we focus in the study of the asymptotic properties of a cosmological model with interesting features, namely a phase of current cosmic acceleration and the absence of an initial singularity, which was proposed in \cite{novellocqg} (see \cite{23}) and is based upon the following Lagrangian density:

\be {\cal L}=-\frac{1}{4}F+\alpha F^2+\gamma F^{-1},\label{L}\ee where $\gamma$ is a new overall (constant) parameter. The second (quadratic) term dominates during very early epochs of the cosmic dynamics, while the Maxwell term (first term above) dominates in the radiation era. The last term in (\ref{L}) is responsible for the accelerated phase of the cosmic evolution \cite{novellocqg}, so that here we can safely ignore the cosmological constant term $\Lambda$ in Einstein's field equations (\ref{feqs}). The above Lagrangian density yields a unified scenario to describe both the acceleration of the universe (for weak fields) and the avoidance of the initial singularity, as a consequence of its properties in the strong-field regime. 

After averaging over a (flat) FRW space-time the stress-energy tensor associated with (\ref{L}) can be written in the form of an equivalent perfect fluid stress-energy tensor with energy density and parametric pressure:

\bea &&\rho_B=\frac{B^2}{2}\left(1-8\alpha B^2-\frac{\gamma}{B^4}\right),\nonumber\\
&&p_B=\frac{B^2}{6}\left(1-40\alpha B^2+\frac{7\gamma}{B^4}\right),\label{rhop}\eea respectively. Notice that, for large values of the NLED field, positivity of energy requires that $B<1/\sqrt{8\alpha}$, while, for small enough values of $B\ll 1$, positivity of energy implies that $B>\gamma^{1/4}$. Therefore, in the unified model of \cite{novellocqg} (Lagrangian density (\ref{L})) there arise both higher and lower bounds on the values the NLED field $B$ can take. However, given that the observational data constraints the parameter $\gamma$ to be $\sqrt{|\gamma|}\approx 4\times 10^{-28}\;g\;cm^{-3}$ \cite{23}, then the lower bound on $B$ can be admitted without going into conflicts with observations. 

Our goal in this section will be to put the cosmological equations (\ref{feqs}), (\ref{contcdm}), and (\ref{cont}), with $\rho_B$ and $p_B$ given by (\ref{rhop}), in the form of an autonomous system of ODE. Recall that, since the NLED Lagrangian density (\ref{L}) accounts also for the present accelerated stage of the cosmic expansion, then in the Einstein's field equation (\ref{feqs}) the cosmological constant is set to zero $\Lambda=0$. In the present case we choose the following phase space variables:

\be x\equiv\frac{B}{\sqrt 6 H},\;v\equiv 4\sqrt\alpha H,\;w\equiv\frac{\sqrt\gamma}{6 H^2}.\label{vars}\ee The corresponding autonomous system of ODE can be written in the following form:

\bea &&x'=-x\left(2+\frac{H'}{H}\right),\nonumber\\
&&v'=v\;\frac{H'}{H},\;w'=-2w\;\frac{H'}{H},\nonumber\\
&&\frac{H'}{H}=-\frac{3}{2}-\left(\frac{x^4-15x^6 v^2+7w^2}{2x^2}\right).\label{ode1}\eea The phase space relevant for this case can be defined as:

\bea &&\Psi_U=\{(x,v,w):0\leq x^4-3x^6v^2-w^2\leq x^2,\nonumber\\&&\;\;\;\;\;\;\;\;\;\;\;\;\;\;\;\;\;\;\;\;\;\;\;\;\;\;\;\;\;\;\;\;\;\;\;x\geq 0,\;v\geq 0,\;w\geq 0\}.\label{phasespace1}\eea

In terms of the above phase space variables:

\bea &&\Omega_B=\frac{x^4-3x^6 v^2-w^2}{x^2},\nonumber\\
&&\omega_B=\frac{1}{3}\left(\frac{x^4-15x^6 v^2+7w^2}{x^4-3x^6 v^2-w^2}\right).\label{params}\eea 

\begin{figure}[t!]
\begin{center}
\includegraphics[width=8cm,height=6.5cm]{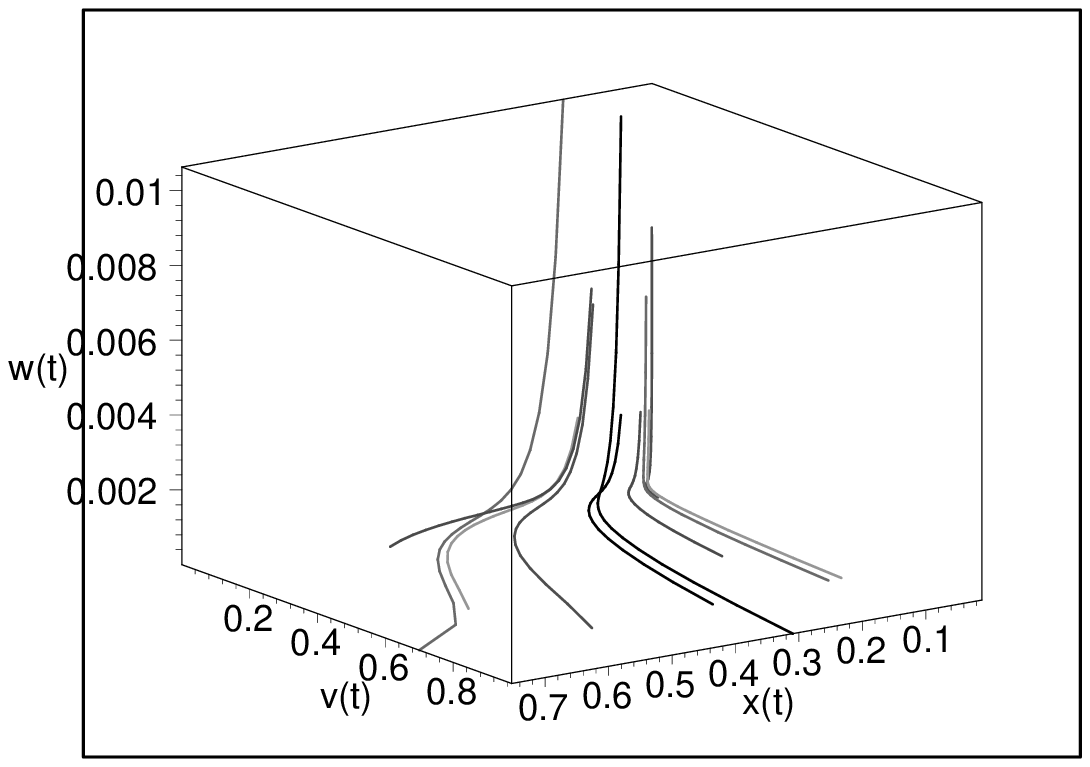}
\includegraphics[width=4cm,height=4cm]{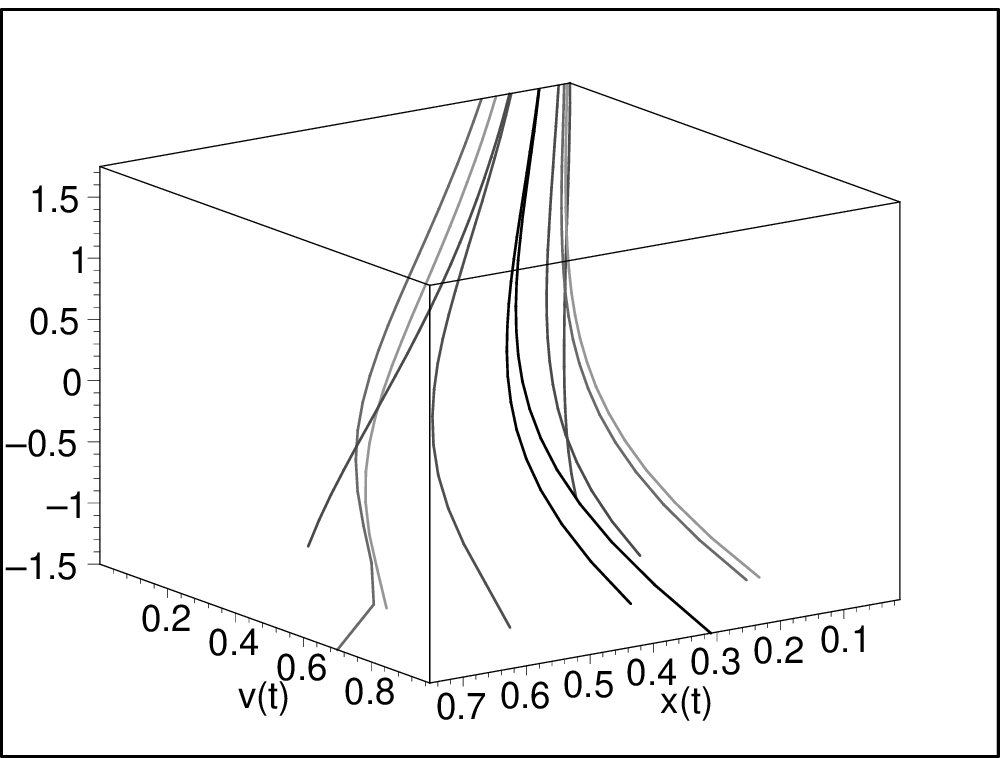}
\includegraphics[width=4cm,height=4cm]{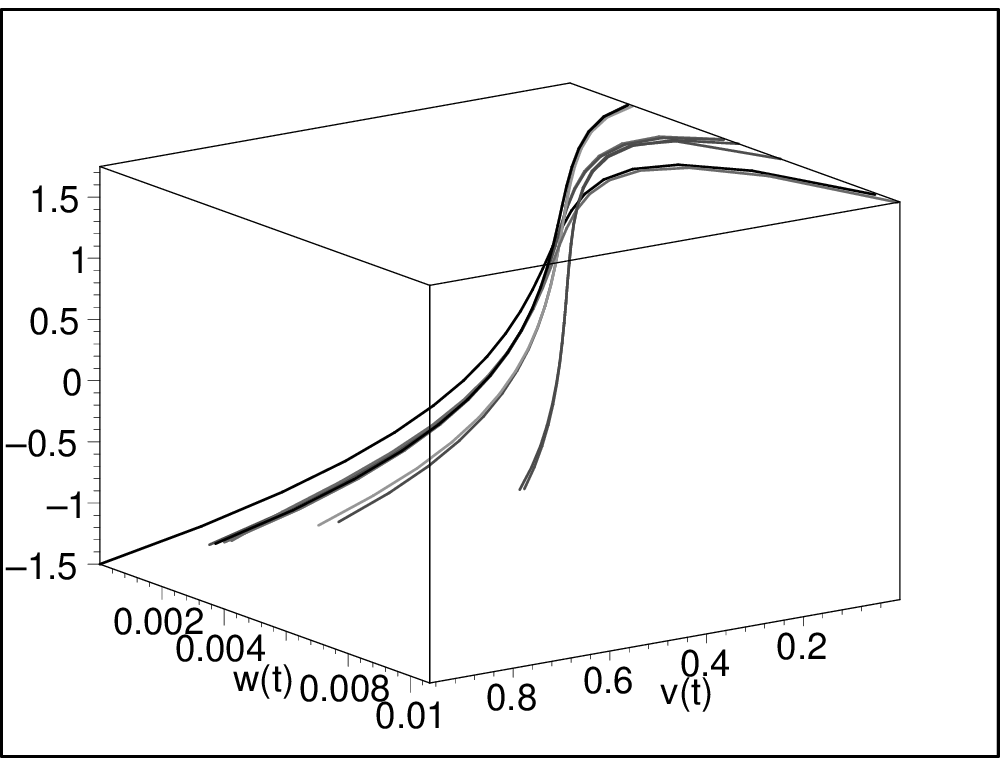}
\vspace{0.3cm}
\caption{The orbits of the autonomous system of ODE (\ref{ode1}) for arbitraryly chosen sets of initial conditions (upper panel). Only expanding cosmologies are reflected in the phase space. Different perspectives of the flux in time of the orbits are shown in the lower panels. Notice that the solution dominated by the CDM (point $(0,0,0)$ in Tab.\ref{tab1'}) is always a saddle point.}\label{fig2}
\end{center}
\end{figure}

There can be found only two critical points of the autonomous system of ODE (\ref{ode1}) -- shown in Tab.\ref{tab1'}: i) the CDM-dominated solution (point $P_1$ in Tab.\ref{tab1'}), and ii) the radiation-dominated solution (critical point $P_2$). As it is apparent from Tab.\ref{tab1'}, both represent saddle equilibrium points in the phase space $\Psi_U$ (\ref{phasespace1}). Notice that there is neither past nor future attractors that could be associated with the model of this section. This fact is specially illustrated in the figure Fig.\ref{fig2} where the orbits of the system of ODE (\ref{ode1}) are plotted in a portion of the 3D phase space $(x,v,w)\in\Psi_U$. 

In the present model, as in the model of section III, the possibility that the NLED field might cause the late-time accelerated expansion to happen -- although existing --, is not a generic property of the model, since there are no critical points that could be correlated neither with early-time nor with late-time NLED effects. This fact is reflected in the figure Fig.\ref{fig2}, where it is apparent that the orbits of the system of ODE evolve from points that are dominated by early-time non-linear effects (large $v$-s and vanishing $w$-s) towards larger values of the variable $w$, that amounts to increasing relevance of the last term in the NLED Lagrangian density ${\cal L}$ of (\ref{L}). However, as clearly seen, there is neither any past (attractor) source points from which the orbits are repelled in the past, nor converging focus towards which the orbits are attracted in the future.

\section{Discussion and Conclusions}

Non-linear electrodynamics can supply a useful scenario where to discuss two relevant problems of the standard cosmological model: i) the initial singularity related with the big-bang, and ii) the present stage of accelerated expansion of the universe. A Lagrangian density of the form given in (\ref{lag}) might provide a solution to the first problem, since in this model the universe undergoes a bounce without singularity of the curvature invariants. There can be also a chance for this model to fuel the early-time inflation. A model based on the Lagrangian (\ref{L}), instead, can account for the present accelerated stage of the cosmic evolution, as well as for avoidance of the initial singularity, in a unified picture. 

A strightforward analysis of the ordinary differential equations (\ref{ode}) or (\ref{ode1}) reveals, however, that there are no critical points in the phase space of the above models that could be associated with non-linear effects. Actually, in both models, existence of equilibrium points correlated with non-linear effects requires that $x'=v'=w'=0$ and, simultaneously $x\neq 0$, $v\neq 0$, and $w\neq 0$.\footnote{The variable $v$ accounts for the influence at early times of the NLED term that is multiplied by the parameter $\alpha$ in (\ref{L}) -- as well as in (\ref{lag}) --, while $w$ is correlated with the term multiplied by $\gamma$ in (\ref{L}) which dominates at late times.} However, a simple inspection of the corresponding equations in (\ref{ode}) and (\ref{ode1}) reveals that the simultaneous fulfillment of the above conditions is imposible.

The knowledge of the equilibrium points in the phase space corresponding to a given cosmological model is a very important information since, independent on the initial conditions chosen, the orbits of the corresponding autonomous system of ODE will always evolve for some time in the neighbourhood of these points. Besides, if the point were a stable attractor, independent on the initial conditions, the orbits will always be attracted towards this point (either into the past or into the future). Going back to the original cosmological model, the existence of the equilibrium points can be correlated with generic cosmological solutions that might really deside the fate and/or the origin of the cosmic evolution. In a sense the knowledge of the asymptotic properties of a given cosmological model is more relevant than the knowledge of a particular analytic solution of the corresponding cosmological equations. While in the later case one might evolve the model from given initial data giving a concrete picture that can be tested against existing observational data, the knowledge of the asymptotic properties of the model gives the power to realize which will be the generic behaviour of the model without solving the Einstein's field equations. In the dynamical systems language, for instance, a given particular solution of the Einstein's equations is just a single point in the phase space. Hence, phase space orbits show the way the model drives the cosmological evolution from one particular solution into another one. Equilibrium points in the phase space will correspond to solutions of the cosmological (Einstein's) equations that, in a sense, are preferred by the model, i. e., are generic. The lack of equilibrium points that could be correlated with a given analytic solution of the model, amounts to say that this solution is not quite generic, otherwise unstable in terms of phase space variables, and can not be taken too seriously.

This is precisely the case with the models under study in the present paper. The analysis of the Einstein's field equations, together with the conservation equations, shows that, in principle, it is possible that the effects of NLED fields might originate the early-time inflation as well as the late-time stage of accelerated expansion of the universe. The possibility to avoid the initial singularity, replacing it by a bounce, seems to be another relevant property of the model. However if one applies the tools of the dynamical systems approach, one can realize that this is no more than just a hope. Actually, solutions where the big-bang singularity is replaced by a bounce, or where the NLED effects fuel primordial inflation and/or late-time accelerated dynamics are not actually preferred by the model. Instead, the solutions dominated by CDM and by standard radiation represent equilibrium points in the phase space, meaning that the cosmic evolution might transit through the corresponding stages during a quite long time. A clear deficiency of the models is the inexistence of past attractors, meaning that the initial conditions from which the universe might be evolved (in the vicinity of the bounce) are not generic neither. The inexistence of late-time attractors in the second model (unified description of bouncing cosmology with late-time accelerated expansion), means that the fate of the cosmic evolution in the model, although inflationary, is quite uncertain.

Summarizing we can say that the second model, where the bounce and the late-time accelerated expansion are described in a unified picture, although at first sight seems to be a good candidate to address the initial singularity and the late-time inflation problems at once, in fact can not be a competitive model since both phenomena are not as generic as expected, so that it suffers from such serious drawbacks as the fine tunning and the coincidence problems. The $\Lambda$-CDM-NLED model, instead, is a good candidate to address the late-time acceleration of the expansion thanks to the $\Lambda$-term. In general the cosmic evolution in this model transits by the radiation-dominated followed by the CDM-dominated (intermediated) stages -- corresponding to equilibrium points in the phase space of the model --, to eventually evolve into the de Sitter solution $3H^2=\Lambda$. However, the behaviour that is determined by the NLED term in this case -- the bounce -- can not be associated with any equilibrium point in the phase space, meaning that, as in the second model, the NLED effects are not generic and can be met just as a particular solution of the Einstein's field equations.

This work was partly supported by CONACyT M\'{e}xico, under grants 49865-F, 54576-F, 56159-F, 49924-J, 105079, 52327, and by grant number I0101/131/07 C-234/07, Instituto Avanzado de Cosmologia (IAC) collaboration. R. G.-S. acknowledges partial support from COFAA-IPN and EDI-IPN grants, and SIP-IPN 20090440. T. G. and I. Q. aknowledge also the MES of Cuba for partial support of the research.

\end{document}